# First-principles study of ultrathin (2×2) Gd nanowires encapsulated in carbon nanotubes

*Jae-Hyeon Parq*,[1] *Jaejun Yu*,[1] *and Gunn Kim*[2,a)]

[1]Department of Physics and Astronomy, Center for Strongly Correlated Materials, Seoul National University, Seoul 151-747, Korea

[2]FPRD and Department of Physics and Astronomy, Seoul National University, Seoul 151-747, Korea

Using density functional calculations, we investigate the structural and magnetic properties of ultrathin Gd and Gd-carbide nanowires (NWs) encapsulated in narrow carbon nanotubes (CNTs). The equilibrium geometry of an encapsulated (2×2) Gd-NW is markedly different from that of bulk Gd crystals. The charge-density analysis shows pronounced spin-dependent electron transfer in the encapsulated Gd-NW in comparison with that of Gd-carbide NWs. We conclude that Gd-CNT hybridization is primarily responsible for both the structural difference and electron transfer in the encapsulated Gd-NW.

---

[a)]Electronic mail: gunnkim@phya.snu.ac.kr (G. Kim)

## I. Introduction

One-dimensional (1D) nanostructures have been extensively studied because of their potential applications in nanoelectronic devices, for the transport of charge, spin, and heat or for optical excitation.[1,2] However, 1D metal nanowires (NWs) are fragile and chemically unstable, and the encapsulation of NWs inside carbon nanotubes (CNTs) has been considered as one of the best solutions to compensate for these drawbacks. The strong $sp^2$ covalent bonding of C atoms in the CNT is responsible for their excellent mechanical properties, and the hollow space in the CNT can be used for the encapsulation of metal NWs. Encapsulation can protect 1D NWs from harm caused by the external environment, such as oxidation or ion bombardment. Further, the geometry of the CNT constrains an encapsulated metal NW by restricting the size of the CNT, e.g., by restricting the diameter of the CNT.

Among the various types of encapsulated metal NWs, transition-metal and rare-earth-metal NWs have attracted relatively greater attention because of their fundamental electronic and magnetic properties as well as their potential for the applications in spintronics.[3–8] Recently, Kitaura *et al.* synthesized Gd nanowires (Gd-NWs) inside CNTs and, by using high-resolution transmission electron microscopy (HRTEM), showed that the encapsulated thin Gd-NWs with very small diameters have geometries which are quite different from those of 1D segments of bulk Gd crystals.[9] Direct measurements of the interatomic distance using HRTEM images showed that the Gd-Gd distance is 0.41 nm, which is larger than the bond length of the bulk Gd crystals. It was suggested that charge transfer between encapsulated Gd atoms and the CNT plays a crucial role in stabilizing the observed Gd-NW structures with large Gd-Gd distances, which cannot be explained by simple rigid-sphere filling.[9] Such charge transfer has been reported for other encapsulated NWs,[10–12] resulting in a reduction in magnetization.

The structure and composition as well as the electronic and magnetic properties of Gd-NWs are yet to be established. Since TEM is not sensitive to light atoms such as C atoms, as Kitaura *et al.* suggested, it is hard to distinguish NWs consisting of pure Gd atoms from those consisting of Gd-carbides in which the C atoms are located at the center of the Gd-carbide NW.[9] In order to get an insight into the observed Gd-NW structures, it is necessary to investigate the structures and possible configurations of encapsulated Gd-NWs. Furthermore, the electronic and magnetic properties of 1D Gd-NWs have not been well understood in comparison with the ferromagnetic metallic state of Gd bulk crystals. To the best of our knowledge, there is no theoretical or experimental report on the magnetic properties of ultrathin Gd-NWs. In this paper, we report first-principles-calculation results for the structural and magnetic properties of rectangular (2×2) Gd and Gd-carbide NWs. By comparing encapsulated Gd and Gd-carbide NWs with free-standing NWs, we show that the hybridization interaction between Gd and the C atoms either at the CNT wall or inside the Gd-carbide NW is crucial in determining the equilibrium structures. From the total-energy calculations for the candidate structures, we confirm that the observed Gd-Gd distance,[9] which is much larger than the Gd-Gd distance in the Gd bulk crystal, is consistent with the corresponding values in the encapsulated Gd-NW in the CNT. We also elucidate the difference in the structure of the encapsulated nanowires between Gd-NW and Gd-carbide NWs.

## II. Computational details

The first-principles calculations were carried out using density-functional-theory (DFT) code, OpenMX,[13] within the local spin density approximation plus Hubbard U (LSDA+U) framework.[14] Our OpenMX code employs a linear combination of pseudo-atomic orbitals (LCPAO) method,[15] where the pseudo-atomic orbitals (PAOs) were chosen as C 4.5-s2p2 for carbon and Gd 9.0-s2p1d1f1 for gadolinium. We used Troullier-Martins-type norm-conserving pseudopotentials[16,17] in a factorized separable form with partial core correction.[18] The effective on-side Coulomb interaction parameter $U_{\mathrm{eff}}$ was 6 eV for the Gd 4f orbitals[19] in the dual-occupation representation.[14] Real-space grid techniques[20] were used with the energy cutoff up to 350 Ry in numerical integration, and the Poisson equation was solved using the fast Fourier transformation (FFT) technique. In our calculations, 30 × 1 × 1 k-points were sampled for primitive cell calculations, and 16 × 1 × 1 k-points for double unit-cell calculations. The model structures were relaxed until the Hellman-Feynman force was less than $6 \times 10^{-4}$ Hartree/Bohr.

To investigate the equilibrium geometries, we started with the model structures of (2×2) Gd-NWs encapsulated in the (14,0) CNT and considered the possible Gd-carbide NWs, as illustrated in Fig. 1. The lattice structure of the (2×2) Gd-NW is defined by the lattice constant *a* along the axial direction of the wire and the lateral Gd-Gd distances *b* and *c* perpendicular to the wire direction, as shown in Fig. 1(a). In the case of the Gd-carbide NW, we considered two possible configurations of C dimers: perpendicular to the axial direction (the "Gd-carbide A" structure) as shown in Fig. 1(b) and parallel to the axial direction (the "Gd-carbide B" structure) as shown in Fig. 1(c). The lattice structures of Gd-carbides A and B are described by the same parameters *a*, *b*, and *c*, respectively. In order to carry out calculations for encapsulated and free-standing NWs, we used the supercell and set the separation between neighboring CNTs or NWs at 20 Å to avoid possible interference between neighboring cells. To study the dependence of encapsulated Gd-NW on the chirality of the CNT, we also calculated Gd-NW@(8,8) CNT.

## III. Results and discussion

To compare the structures of various NWs *with and without* encapsulation, we first calculated the relaxed geometries and magnetic moments of different free-standing NWs without encapsulation. The optimized lattice structures and magnetic moments of the free-standing NWs are listed in Table I. The calculated lattice constant, *a* = 3.18 Å, of the free-standing Gd-NW is found to be much smaller than the Gd-Gd distances in Gd bulk crystals, which are 3.57 Å and 3.5 Å for hcp and bcc structures, respectively. The lateral Gd-Gd distances *b* and *c* of the free-standing Gd-NWs are even smaller than those in the Gd bulk crystals. On the other hand, the lattice constants of Gd-carbides A and B are not equal even though the only structural difference is the orientation of the C dimer with respect to the axial direction of the wire. While the lattice constant, $a$ = 3.12 Å, of the free-standing Gd-carbide A is close to that of the free-standing Gd-NW, the lattice constant, $a$ = 3.75 Å, of Gd-carbide B is comparable to the Gd-Gd distance in tetragonal Gd-carbide bulk crystals. On the basis of the observation that the Gd-C distances are 2.41 Å and 2.43 Å for Gd-carbides A and B, respectively, it can be assumed that Gd-C bonding plays a crucial role in determining the structures of Gd-carbide NWs. From the total energy calculations, the energy of Gd-carbide A is lower than that of Gd-carbide B by 0.23 eV/Gd-atom.

The lateral Gd-Gd distances listed in Table I exhibit an interesting pattern for different configurations of the Gd and Gd-carbide NWs. For the pure Gd-NW, $b$ and $c$ are relatively smaller than the lattice constant $a$, but they increase in the case of Gd-carbide A. The changes in $b$ and $c$ of Gd-carbide B are opposite to those of Gd-carbide A. While the reduced distance in the free-standing Gd-NW can be understood by the loss of neighboring atoms along the lateral directions, the lattice constant and Gd-Gd distances for Gd-carbides A and B seem to be more dependent on the Gd-C bonding structures, e.g., the orientation of the C dimers. These results demonstrate that the position and orientation of C dimers are critical in determining the overall structure of free-standing Gd-carbide NWs and further emphasize the importance of Gd-C bonding, which is much stronger than Gd-Gd bonding.

TABLE I. Gd-Gd distances ($a$, $b$, $c$), Gd-C (dimer) distances, and magnetic moments per Gd atom of free-standing NWs.

| | $a$ (Å) | $b$ (Å) | $c$ (Å) | Gd-C (Å) | Mag. Mom./Gd ($\mu_B$) |
|---|---|---|---|---|---|
| Gd | 3.18 | 3.09 | 3.09 | - | 8.11 |
| Gd-carbide A | 3.12 | 3.18 | 3.32 | 2.41 | 7.94 |
| Gd-carbide B | 3.75 | 3.00 | 3.00 | 2.43 | 8.07 |

TABLE II. Gd-Gd distances ($b$, $c$), Gd-C (dimer) distances, separations from Gd to CNT wall and magnetic moments per Gd atom of Gd-NWs@(14,0)CNT at the fixed lattice constant $a$ = 4.29 Å.

| | $b$ (Å) | $c$ (Å) | Gd-C (Å) | Gd-CNT (Å) | Mag. Mom./Gd ($\mu_B$) |
|---|---|---|---|---|---|
| Gd | 4.71 | 4.15 | | 2.28 | 8.05 |
| Gd-carbide A | 2.94 | 3.24 | 2.75 | 3.23 | 8.12 |
| Gd-carbide B | 2.96 | 2.97 | 2.52 | 3.38 | 7.83 |

Since the Gd-C interaction plays an important role in determining the configuration of the free-standing Gd-carbide NWs, it is expected that the same Gd-C interaction play a similar role in determining the structure of the encapsulated Gd and Gd-carbide NWs in the CNT, as it does in the free-standing Gd-carbide NWs. To investigate the interaction of Gd atoms with the C atoms in the CNT wall, we examined the binding characteristics of a Gd atom adsorbed on the inner wall of the (14,0) CNT. Gd atoms at three different sites of (a) an on-top, (b) a bridge, and (c) a hollow configuration are shown in Fig. 2. At each site, the minimum-energy position of the Gd atoms was determined by varying the distance between the Gd atom and the CNT wall. The binding energies of the on-top, bridge, and hollow sites are found to be 1.45, 1.55, and 1.91 eV, respectively. The distance between the Gd atom and the CNT wall was 2.22 Å at the hollow site, while the corresponding distances at the on-top and bridge sites were 2.34 Å and 2.30 Å, respectively. Here, it is noted that the actual Gd-C distance at the hollow site is larger than the Gd-CNT wall distance. Gd-C distances of Gd atoms adsorbed on CNTs is comparable to those of both Gd-carbides A and B listed in Table I. The strong binding of a Gd atom at the hollow site of the CNT implies that the Gd-CNT interaction should be a major factor in determining the structure of encapsulated Gd-NWs.

The minimum energy configuration of the (2×2) Gd-NW by performing structural relaxation for a rectangular (2×2) Gd-NW inside the (14,0) CNT. Table II lists the Gd-Gd distances for the encapsulated Gd-NW in the CNT (Gd-NW@CNT) for a fixed lattice constant $a$ = 4.29 Å, which matches the lattice period of CNT. We performed a structural relaxation calculation for the Gd-NW@CNT and found the Gd-NW lattice constant of ~ 4.3 Å is comparable to the lattice period of the (14,0) CNT. The values of *b* and *c* of the Gd-NW@CNT are 4.71 Å and 4.15 Å, respectively, and they are much larger than those for the bulk Gd crystals. On the other hand, the distance between the Gd atoms and the CNT wall is 2.28 Å, which is close to that for a single Gd atom at the hollow site of the CNT. As a Gd atom preferentially occupies the hollow sites of CNT, the Gd atoms in Gd-NW@CNT prefer the hollow-site configuration so as to attain the minimum-energy configuration. A large increase in the Gd-Gd distance with a slight change in the Gd-CNT distance indicates that the Gd-CNT interaction is much stronger than the Gd-Gd bonding in the NW itself. Thus, the structure of the encapsulated Gd-NW in the CNT is significantly different from that of its bulk counterpart. Indeed, the calculated Gd-Gd distances in the encapsulated Gd-NW are in good agreement with experiment.[9] The difference is related to the strong interaction between the Gd atoms and the CNT, which is likely to occur only in narrow Gd-NWs and is likely to weaken in thick Gd-NWs where the Gd-Gd distances will return to the bulk values.

To investigated the electronic and magnetic properties of all three encapsulated NW models: pure Gd, Gd-carbide A, and Gd-carbide B, we carried out the structural relaxation calculations. For the sake of comparison of Gd-NW@CNT with Gd-carbide NW@CNT, we fixed the lattice constant of the encapsulated Gd-carbide A and Gd-carbide B NWs at that of Gd-NW@CNT. While the commensurate structure of Gd-NW@CNT is a consequence of the dominant Gd-CNT interaction, the lattice structure of Gd-carbide NW@CNT is more subtle. Because of the competition between the Gd-C interaction within the Gd-carbide NWs and the Gd-CNT interaction, the Gd-carbide NW@CNT may have an incommensurate lattice structure. The energetics in Gd-carbide NWs seems to be associated with the increased distance between the Gd and C atoms, as listed in Table II, and the distorted CNT structure of Gd-carbide A (see Fig. 1). The Gd-carbide B structure is energetically favored over the Gd-carbide A structure by 0.36 eV/Gd-atom. In contrast to the strong Gd-CNT bonding in the case of pure Gd-NW, the Gd-CNT wall distances for Gd-carbides A and B change to 3.23 Å and 3.38 Å, respectively. Further, the lateral Gd-Gd distances, i.e., *b* and *c*, are still close to those of the free-standing NWs. The weakened Gd-CNT interaction and the invariance of the lateral Gd-Gd distances reflect the strength of the Gd-C interaction in Gd-carbides.

Recent experiments[9,21] have reported charge transfer between the Gd-NW and the CNT. To demonstrate the effect of charge transfer, we plotted the charge density difference between an isolated CNT plus an isolated Gd-NW (or a Gd-carbide NW) and the encapsulated Gd-NW@CNT. Figure 3 shows the charge-density difference plots for the encapsulated (a) Gd-NW and (b) Gd-carbide A NWs. The case of Gd-carbide B is similar to that of Gd-carbide A. Figure 3(a) shows a large charge transfer from the Gd-NW to the CNT; this occurs because of the strong Gd-CNT interaction. On the other hand, the suppressed Gd-CNT interaction in the case of Gd-carbide A is indicated in Fig. 3(b), where little charge transfer occurs between Gd and CNT.

The difference in the charge-transfer interactions among Gd, Gd-carbide A, and Gd-carbide B can be understood by considering the change of the projected density-of-states (pDOS) at the C atoms in the CNT. The CNT-projected pDOSs for Gd-carbides A and B exhibit an upshift of the Fermi level *without a substantial change* in the electronic structure of the host CNT. On the other hand, the CNT-pDOS of the pure Gd-NW in Fig. 4 shows a significant variation of pDOS, which indicates a significant change in the electronic structure of the host CNT. In the Gd-NW@CNT, the original bands of the (14,0) CNT are downshifted by ~1.3 eV because of the electron transfer from the Gd atoms to the nanotube (Fig. 4). However, the band shifts of the encapsulated Gd-carbide NWs are in the range of only 0.6–0.7 eV. This implies that the Gd-NW @CNT contributes more charge to the CNT than the Gd-carbide NW@CNT. In Fig. 4, the blue dotted lines show the shifted DOS of the bare (14,0) CNT. The primary cause of such change is the strong Gd-CNT hybridization, which makes the Gd 5d band wider than that of the Gd-NW *without* CNT and is responsible for significant charge transfer. The difference between the pure Gd-NW and the Gd-carbide arises from the different degree of Gd-C hybridization.

To elucidate the effect of hybridization in the band structure of Gd-NW@CNT, we show the band structures of Gd-NW@CNT together with the original CNT and the Gd-NW without CNT in Fig. 5. The band structure of Gd-NW@CNT in Fig. 5(c) exhibits several distinct features in comparison with those of the bare CNT and the Gd-NW without the CNT in Figs. 5(a) and (b), respectively. The bandwidth of d-bands of the Gd-NW becomes wide because of the hybridization between Gd and CNT. In addition, the interaction between the Gd-NW and the CNT gives rise to the change in the dispersion shapes of some bands. For instance, near -1.5 eV, a band has a minimum position at $k \neq 0$, as indicated by an arrow in Fig. 5(c). On the other hand, the original band of the corresponding state in the Gd-NW in Fig. 5(b) has the minimum position at Γ-point. Thus, the difference between the band structures shown in Figs. 5(c) and (d) also substantiates the strong Gd-CNT interaction in Gd-NW@CNT.

According to our Voronoi population analysis,[22] it is found that about one electron per Gd atom is transferred from the Gd-NW to the CNT. It is interesting to note that the number of electrons transferred for the majority spins is different from that for the minority spins. The magnetic moment of the free-standing Gd-NW is obtained as 8.11 $\mu_B$ per Gd atom, which is quite close to the bulk value (7.9 $\mu_B$) despite the considerable difference in the lattice constant. The magnetic moments of the free-standing Gd-carbide A NWs are slightly smaller than the magnetic moment of the free-standing Gd-NW. Electron transfer can change the magnetic moment. In fact, the encapsulation reduces the spin moment of the Gd atom by 0.5 $\mu_B$ relative to the spin moment of the Gd atom (8.57 $\mu_B$) in a free-standing (2×2) Gd-NW with the same atomic positions as those in the encapsulated NW (lattice constant: 4.29 Å). In the Gd-carbide NWs, the decrease in the moment per Gd atom is less than 0.1 $\mu_B$ since the electron transfer between the CNT and the Gd-carbide NW is small. As shown in Fig. 4, the DOS of the CNT exhibits a spin polarization because the number of transferred electrons with majority spin differs from that of transferred electrons with minority spin; the transferred electrons are mainly *d* electrons of Gd atoms. It is clearly shown that the DOSs for the majority and minority spins are different in the band gap of the CNT that is used for the encapsulation of the Gd and the Gd-carbide A NWs. For the encapsulated Gd-carbide A and B NWs, ~0.5*e*/Gd-atom is transferred to the host (14,0) CNT. This is because the bonding of Gd atoms with C atoms in the Gd-carbide wire is strong, and the distance between the Gd atoms and the wall of the CNT is relatively large.

TABLE III. Relative energies (E) per Gd atom, Gd-Gd distances (*a*, *b*), separations from Gd to CNT wall and the twisting angle (θ) of Gd rectangles with respect to the adjacent Gd rectangles in Gd-NWs@(8,8) CNT with a lattice periodicity of 7.43 Å. Here, FM and AFM represent ferromagnetic and antiferromagnetic configurations, respectively.

| | | E (meV/Gd) | *a* (Å) | *b* (Å) | Gd-CNT (Å) | θ (°) |
|---|---|---|---|---|---|---|
| Armchair A | FM | 162 | 3.74 | 4.38, 4.03 | 2.29, 2.51 | 6.6 |
| | AFM | 166 | 3.74 | 4.38, 4.04 | 2.30, 2.50 | 6.6 |
| Armchair B | FM | 4 | 3.96 | 4.37, 4.28 | 2.32, 2.36 | 25.7 |
| | AFM | 0 | 3.98 | 4.38, 4.29 | 2.31, 2.39 | 25.8 |

TABLE IV. Relative energies (E), Gd-Gd distances (*a*, *b*), separations from Gd to CNT wall and Gd-NW twisting angle (θ) of Gd rectangles with respect to the adjacent Gd rectangles in Gd-NWs@(8,8) CNT with a lattice periodicity of 9.91 Å.

| | | E (meV/Gd) | *a* (Å) | *b* (Å) | Gd-CNT (Å) | θ (°) |
|---|---|---|---|---|---|---|
| Armchair C | FM | 0 | 3.35, 6.96 | 4.38, 4.34 | 2.34, 2.35 | 24.9 |
| | AFM | 34 | 3.45, 6.81 | 4.39, 4.37 | 2.33, 2.34 | 23.3 |
| Armchair D | FM | 459 | 4.91, 4.99 | 4.47 | 2.28 | 0.0 |
| | AFM | 412 | 4.91, 5.00 | 4.47 | 2.27 | 0.0 |

So far, we have discussed the structural and electronic properties of the encapsulated Gd and Gd-carbide NWs, with the Gd atoms in a ferromagnetic spin configuration. Although bulk Gd is a ferromagnetic metal, the 1D structure of the NWs may lead to the possible instability of different magnetic configurations. Among the possible magnetic configurations, it is found that the lowest-energy configuration is an antiferromagnetic state with a double unit-cell along the wire direction, where the double-unit cell consists of alternating ferromagnetic layers of (2×2) Gd atoms. The total energy of the ferromagnetic state is higher than that of the antiferromagnetic ground state by 40 meV/Gd-atom and 6 meV/Gd-atom in the free-standing Gd-NW and the encapsulated Gd-NW, respectively. Other spin configurations are rather unstable relative to the both ferromagnetic and antiferromagnetic states. Since the energy differences between the ferromagnetic and antiferromagnetic states are so small, however, a paramagnetic state is expected to prevail in the encapsulated Gd-NWs at room temperatures. The small exchange energy may be due to the large Gd-Gd separation, as in the encapsulated $ErCl_3$ NW.[23] Furthermore, the electronic structures of the two magnetic ground states are almost indistinguishable.

To check the possible alternative structures of the encapsulated Gd-NW in the armchair CNT, we considered the Gd-NW in the (8,8) CNT whose thickness is similar to the (14,0) CNT. A minimal supercell size of the (8,8) CNT is comparable with the Gd-NW in the (14,0) CNT, and consists of 8 Gd atoms within 3 unit cells of the (8,8) CNT (periodicity ≈ 7.43 Å). In this minimal supercell, two possible configurations of Gd atoms are either a nearly rectangular wire (Armchair A) or a twisted wire (Armchair B), as shown in Figs. 6(a) and (b). Ferromagnetic and the antiferromagnetic configurations were also considered as in the zigzag (14,0) CNT. Regardless of the choice of a starting configuration for the structural relaxation, we obtained the optimized structure of minimum energy configuration where the Gd atoms in the (8,8) CNT are localized close to the hollow site of CNT. This result is consistent with that of the Gd-NW in the zigzag CNT where the Gd-CNT interaction dominates over the Gd-Gd interaction. Consequently the relaxed geometry of Gd-NW in the armchair CNT becomes deformed relative to that of the zigzag CNT. Energetically a Gd-NW in the (8,8) CNT cannot have a straight wire geometry regardless of their lattice periodicity because the Gd atoms prefer the hollow sites. It is noted that the hollow site geometry is not compatible with the structure of a straight (2×2) wire. Armchair B has a lower energy (by ~ 160 meV) than Armchair A. As listed in Table III, the Gd rectangles in Armchair B are rotated by ~ 26° alternately with respect to the adjacent Gd rectangles. The twisted form is stable because of the closer position of Gd atoms to the hollow sites. The twisting angle of ~ 26° is associated with the relative positions of hollow sites which are rotated by 22.5° along the tube axis in the armchair CNT. On the other hand, however, the energetically unstable Armchair A Gd NW has a small distortion angle (~ 7° ). When the Gd-CNT distances are relatively larger than those of Armchair B, the Gd atoms of Armchair A may not stick to the hollow site. Thus it is concluded that the Gd-CNT hybridization is again a major factor for the stability of Gd-NW in the armchair CNT. Due to the deformation in the structure of Armchair B, the Gd-Gd distance in the axial direction increases to 3.98 Å, which is definitely larger than that of bulk hcp Gd and considerably larger than those of the free-standing Gd-NWs. The lateral Gd-Gd distances (parameter *b* in Table III) are comparable to the (14,0) CNT encapsulation case (parameter *b* and *c* in Table II).

Since the Gd-Gd distance of 3.98 Å is smaller than the axial Gd-Gd bond length in the zigzag CNT, it is necessary to check whether the larger axial Gd-Gd bond length is possible inside the armchair CNT. To verify this idea, we took a larger supercell of the (8,8) CNT (periodicity ≈ 9.91 Å), and obtained two minimum-energy configurations: Armchair C and Armchair D (shown in Figs. 6 (c) and (d)). In both cases, all Gd atoms are found to be placed near the hollow sites. Therefore, the Gd-CNT distances are maintained close to the other cases, as listed in Table IV. Armchair C is similar to Armchair B, and Armchair D corresponds to a straight rectangular wire. The relative stability of Armchair C to Armchair D is consistent with the stable configuration of Armchair B. Hence, the most probable Gd-NW, inside the armchair (8,8) CNT, has the aligned form of rotated rectangles and the axial Gd-Gd bond length of about 4 Å. For the other chiral CNTs, the encapsulated (2×2) Gd-NWs are expected to have a screw shape corresponding to those of CNTs. The magnetic ground state of the Gd-NW in the CNT may depend on the chirality of the host CNT.

## IV. Conclusion

In summary, we have studied the geometric, electronic, and magnetic properties of Gd and Gd-carbide NWs inside the CNTs to identify the actual features of the synthesized narrow Gd-NW. In the case of the encapsulated Gd-NW, the interaction between the Gd atoms and the CNT wall dominates and results in a large Gd-Gd distance. This is because the Gd-CNT interaction is stronger than the Gd-Gd bonding of the NW itself. When the diameter of the host CNT increases, the encapsulated Gd-NW becomes thicker and the number of neighboring Gd atoms corresponding to a Gd atom increases. In this case, Gd-Gd bonding may act as a major factor in determining the structure of the NW, because of which the Gd-Gd bond length becomes close to that in bulk Gd crystals.[9] In addition, the Gd atoms donate electrons to the host CNT. For the encapsulated Gd-carbide NW, the Gd-Gd separation perpendicular to the tube axis approaches that in free-standing Gd-NWs; this is because of the relatively weaker interaction between the Gd atoms and the CNT. We can conclude that the structured (2×2) Gd-NW presented by Kitaura *et al.*[9] is a pure Gd-NW@CNT. Our pDOS results, which could be measured by scanning tunneling spectroscopy measurements, reveal that the encapsulated Gd-carbide NW might be distinguishable from the encapsulated Gd-NW. It appears difficult to obtain ferromagnetically aligned Gd-NWs that have the rectangular structure even though Gd atoms have large spin moments.

## ACKNOWLEDGMENT

We thank M. J. Han for help in calculation of Gd materials. This work was supported by the second BK21 program (G.K.) and the KOSEF through the ARP (R17-2008-033-01000-0) (J.Y.). We acknowledge the support of KISTI Supercomputing Center under the Supercomputing Application Support Program.

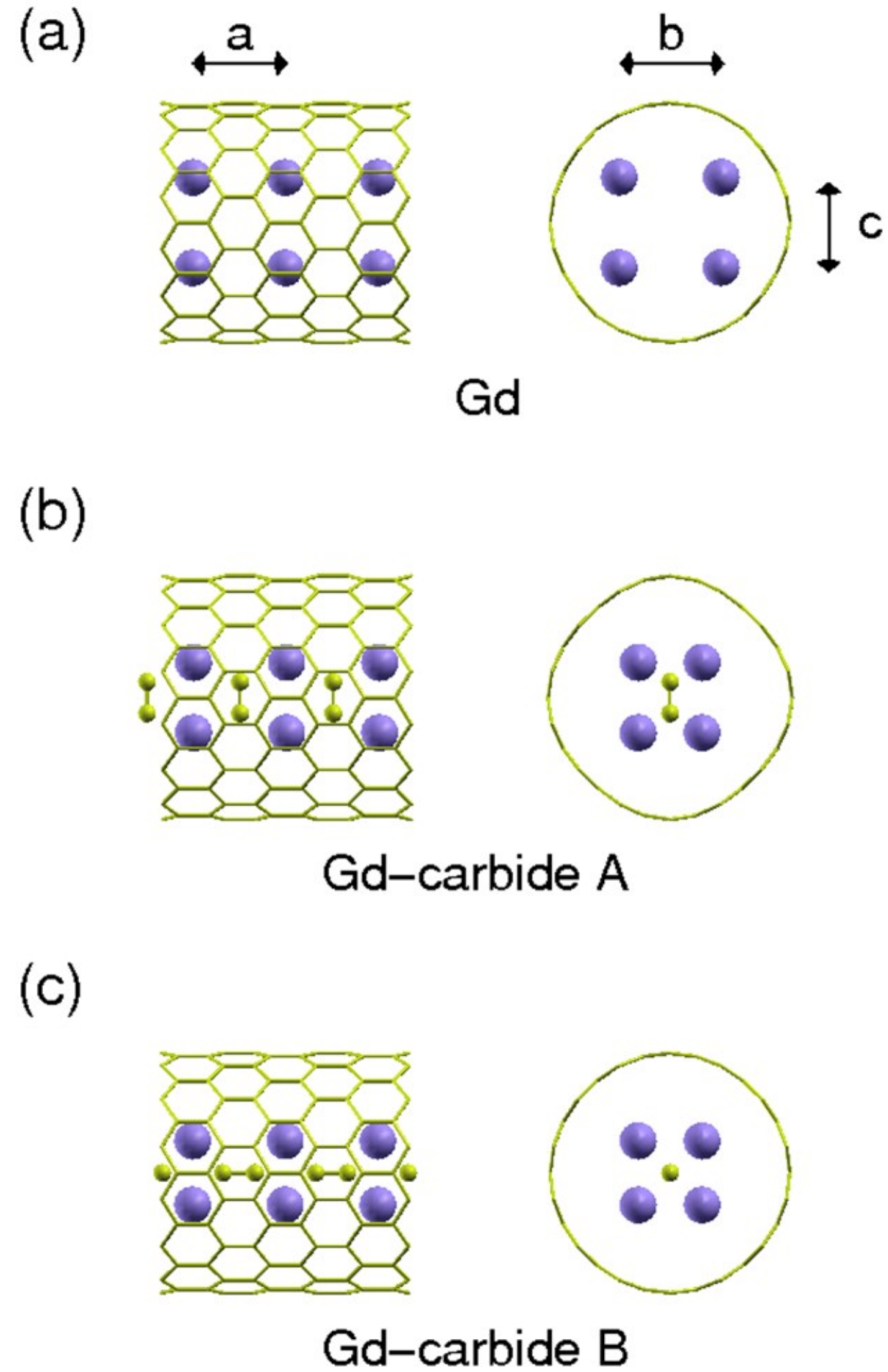

FIG. 1. (Color online) Side view and front view of (a) a model of an encapsulated Gd-NW, (b) a model of an encapsulated Gd-carbide NW that contains perpendicular C dimers and (c) a model of an encapsulated Gd-carbide NW that contains axial C dimers. The Gd-Gd distances are denoted *a*, *b*, and *c*. The structures shown were geometrically optimized.

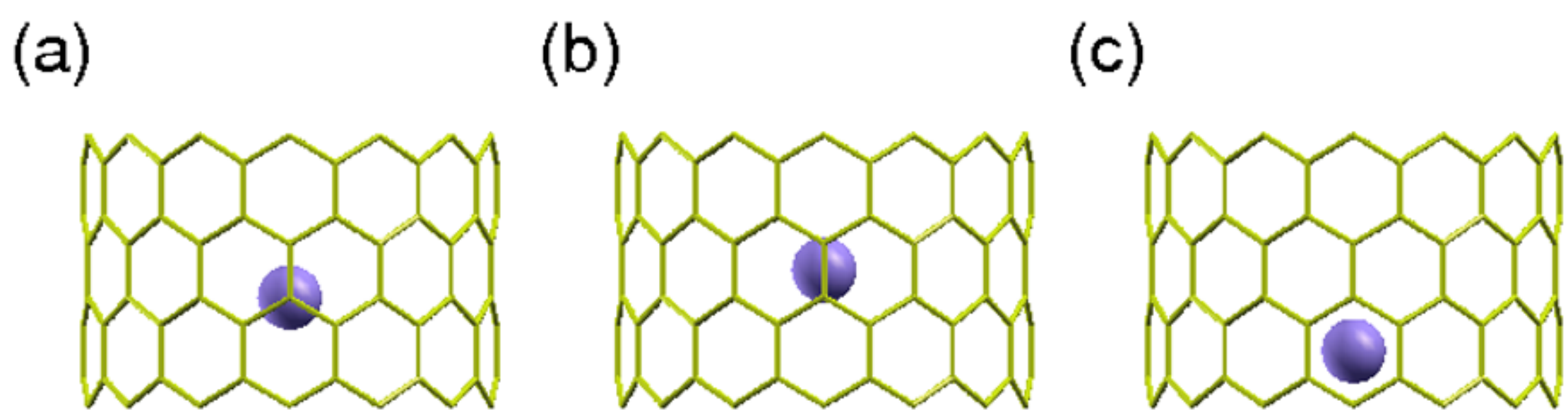

FIG. 2. (Color online) One Gd atom adsorbed on the inner side of the (14, 0) CNT wall. According to the relative positions of the Gd atom, we considered three configurations by: (a) the on-top, (b) the bridge, and (c) the hollow sites.

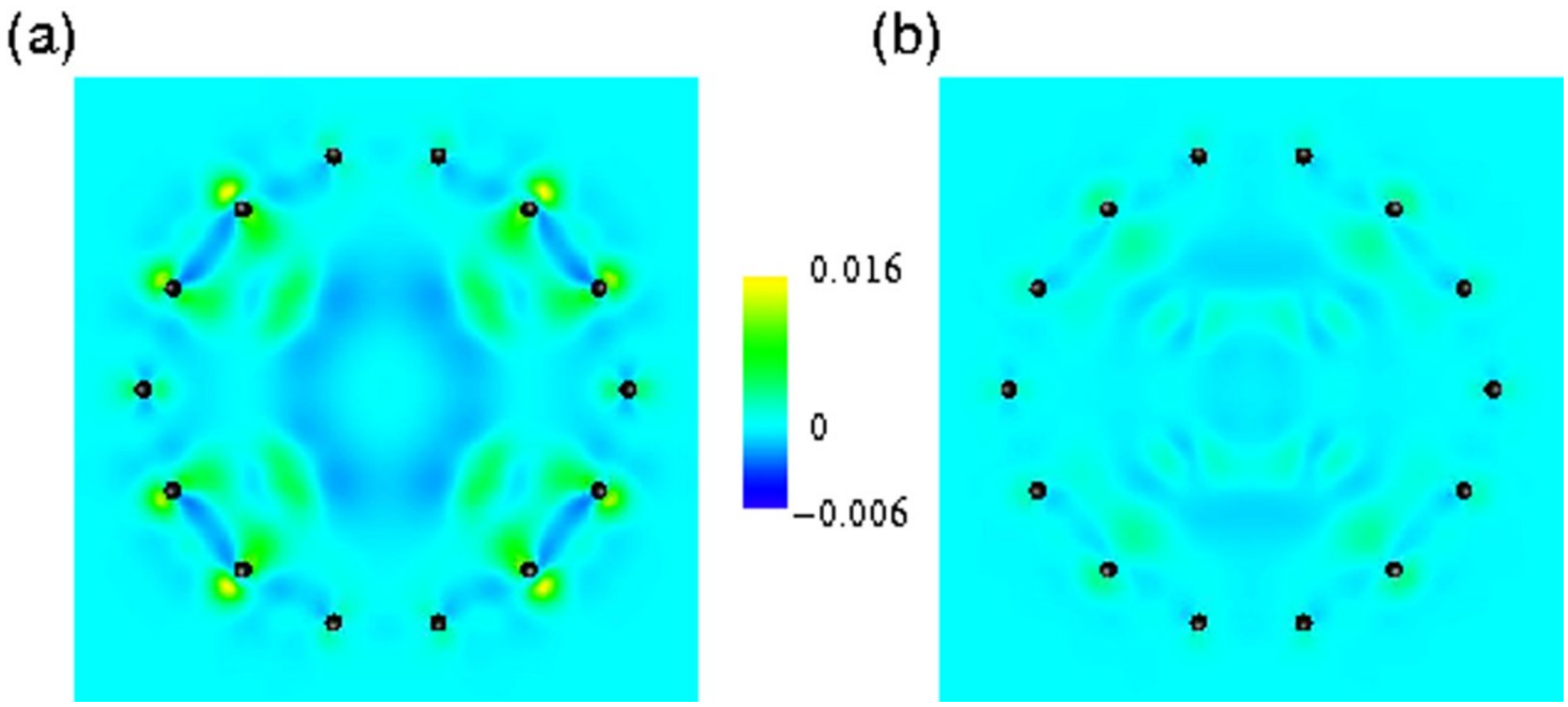

FIG. 3. (Color online) The charge-density differences due to encapsulation for (a) the Gd-NW and (b) the Gd-carbide A NW. The cross section includes C atoms (solid circles).

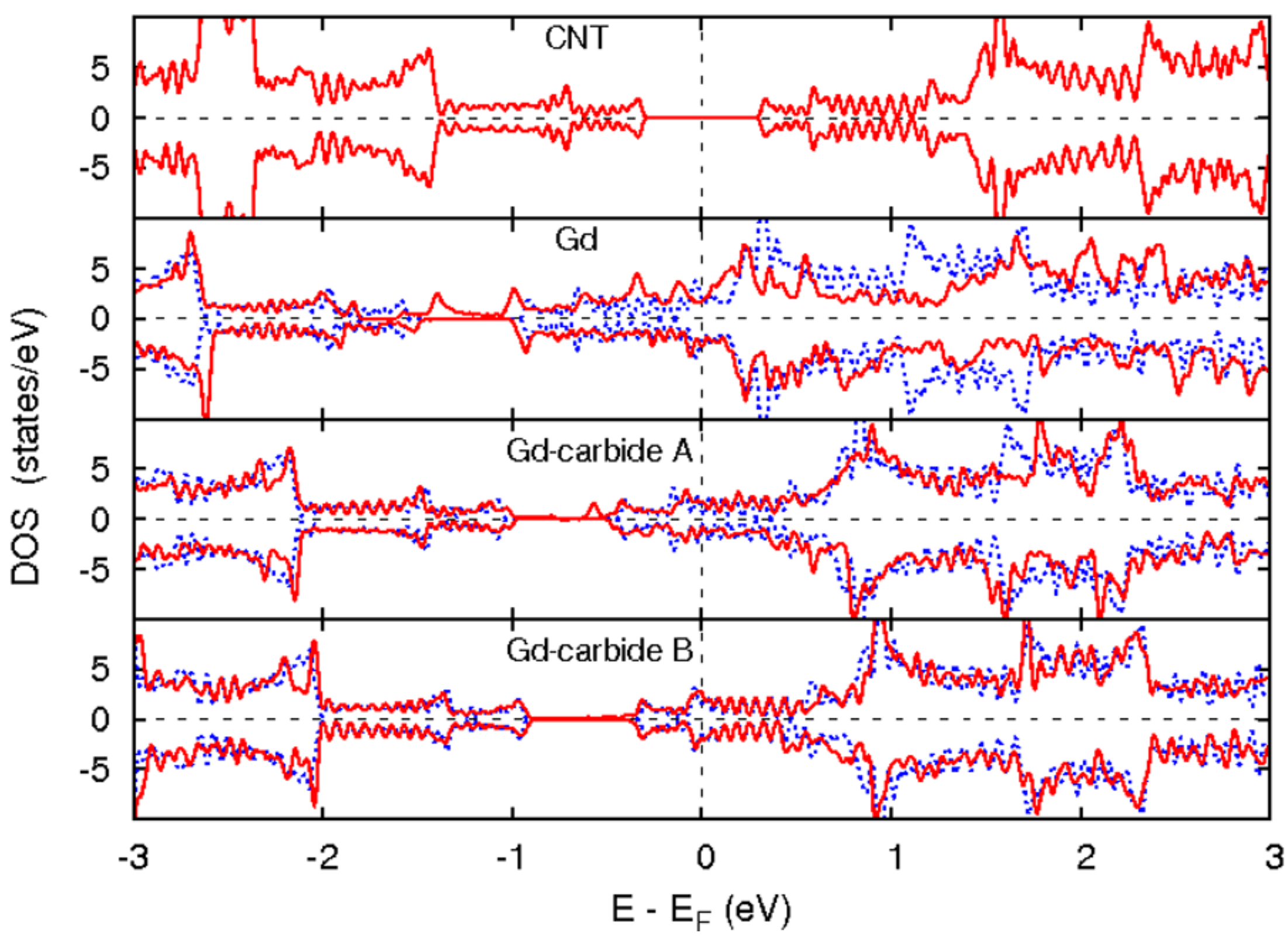

FIG. 4. (Color online) The DOS of the (14,0) CNT (top panel) and the pDOS of the CNT encapsulating the (2×2) Gd-NW (second panel), encapsulated Gd-carbide A (third panel) NW, and Gd-carbie B (bottom panel) NW. The dashed lines represent the shifted DOSs of the bare (14,0) CNT. The positive and negative densities represent major and minor spins, respectively.

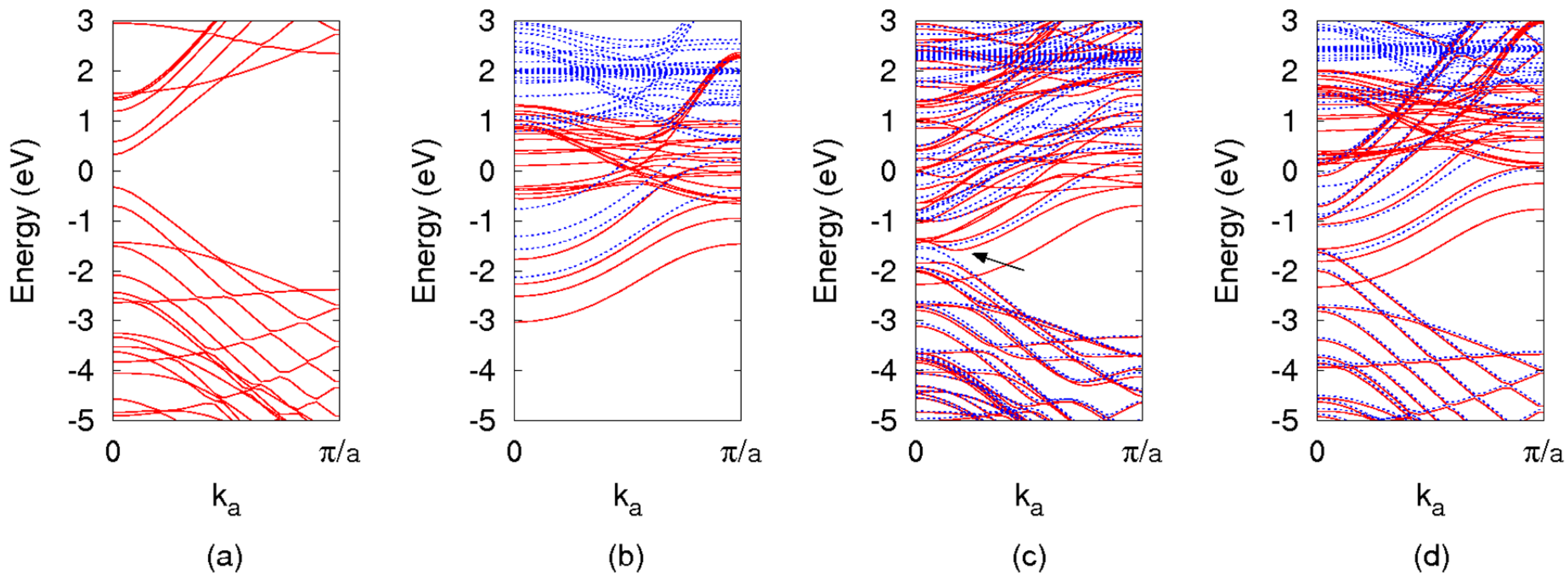

FIG. 5. (Color online) Band structures of (a) the bare (14,0) CNT , (b) the Gd-NW without the (14,0) CNT and (c) the Gd-NW@(14,0)CNT. The Gd-NW without the (14,0) CNT in (b) assumes the position of the Gd atoms in the Gd-NW@(14,0)CNT. (d) is plotted from the superposition of (a) and (b) with a relative shift of the energy bands by 1.8 eV combined with 0.3 eV spin-splitting of Gd d bands to be compared with that in (c). An arrow in (c) indicates a prominent change in the dispersion due to the hybridization between the (2×2) Gd-NW and the (14,0) CNT.

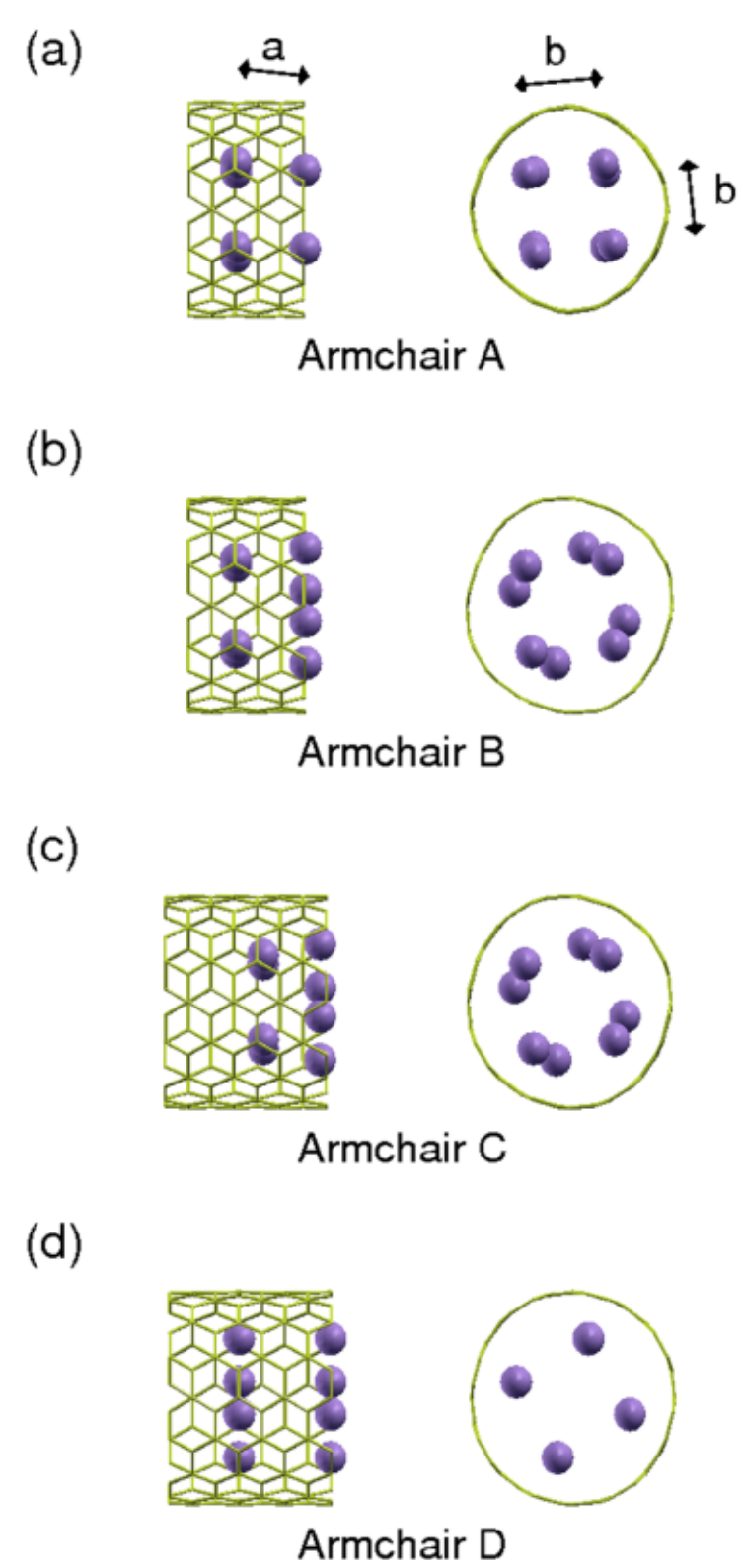

FIG. 6. (Color online) Model structures for the calculated relaxation geometries of the Gd-NW@(8,8)CNT. Index *a* denotes the Gd-Gd bond length along the tube axis while *b* denotes that across the tube axis. Note that Gd rectangles perpendicular to each tube axis are not of the same size, i.e. *b*'s are two values for each Gd-NW except Armchair D.